\documentclass[12pt]{iopart}

\usepackage[numbers]{natbib}

\expandafter\let\csname equation*\endcsname\relax
\expandafter\let\csname endequation*\endcsname\relax
\usepackage{amsmath}

\usepackage{bm}
\usepackage{aas_macros}
\usepackage{graphicx}
\usepackage{pythonhighlight}

\usepackage{iopams}  
\begin{document}

\title[Waveform uncertainty quantification]{Waveform uncertainty quantification and interpretation for gravitational-wave astronomy}

\author{Jocelyn Read}

\address{Nicholas and Lee Begovich Center for Gravitational Wave Physics and Astronomy, California State University Fullerton, Fullerton CA 92831 USA}
\ead{jread@fullerton.edu}

\begin{abstract}
We demonstrate how to quantify the frequency-domain amplitude and phase accuracy of waveform models, $\delta A$ and $\delta \phi$, in a form that could be marginalized over in gravitational-wave inference using techniques currently applied for quantifying calibration uncertainty. For concreteness, waveform uncertainties affecting neutron-star inspiral measurements are considered, and post-hoc error estimates from a variety of waveform models are made by comparing time-domain and frequency-domain analytic models with multiple-resolution numerical simulations. 
These waveform uncertainty estimates can be compared to GW170817 calibration envelopes or to Advanced LIGO and Virgo calibration goals. Signal-specific calibration and waveform uncertainties are compared to statistical fluctuations in gravitational-wave observatories, giving frequency-dependent modeling requirements for detectors such as Advanced LIGO Plus, Cosmic Explorer, or Einstein Telescope.  Finally, the distribution of waveform error for the GW170817 posterior is computed from tidal models and compared to the constraints on $\delta \phi$ or $\delta A$ from GWTC-1 by Edelman et. al. In general, $\delta \phi$ and $\delta A$ can also be interpreted in terms of unmodeled astrophysical energy transfer within or from the source system. 
\end{abstract}

\section{Introduction}

Waveform models are critical for estimating the properties of source systems in gravitational-wave astronomy. In particular for neutron-star mergers,  modeling uncertainties will be potentially significant for tidal inference in coming observation runs \cite{Kunert_2022, Gamba_2021,HuangHasteret2021StatSysNSBH}.  
In the analysis of gravitational-wave observations, these waveform systematics have so far been assessed by comparing or combining inference results from different waveform model families. Examples include  the first binary-black-hole merger GW150914 \cite{2017LVKBBHSyst} and the first neutron-star merger GW170817 \cite{2019LVKGW170817Properties}.  Model selection can also be applied to the predictions of variant waveform families, and marginalization through a mixture model allows the gravitational-wave observation to inform multi-waveform inference \cite{AshtonDietrich2022, 2020AshtonKhanMulti,JanYelikar2020RIFTWF}. 

These methods generally take each waveform family's prediction as an exact model at each point in parameter space. They do not account for modeling uncertainty within each waveform approximant, or its dependence on frequency. Here, we focus on quantifying uncertainty from waveform model choices in the signal domain, in a form that would allow marginalization over the full range of predictions compatible with current theory and simulation. Observed strain data could then inform the best-fit waveform model frequency-by-frequency.
Waveform uncertainty estimates in the signal domain can also be used to guide analysis efforts by providing an indication of how accurately the source will need to be modeled to interpret future signals.

In this work, we present a frequency-dependent characterization of waveform error in amplitude and phase, $\delta A$ and $\delta \phi$,  and demonstrate its application to existing waveform models of neutron star inspirals with tidal contributions. To demonstrate current levels of uncertainty in modeling the physics of waveform generation, we compare multiple waveform approximants with each other and with a high-resolution numerical simulation from the \texttt{CoreDB} waveform library in Section \ref{sec:currentmodels}. The model differences are compared to reference noise spectral densities, anticipating the requirements of future observations. We demonstrate a frequency-dependent condition for the effect of waveform modeling uncertainty to be smaller than statistical fluctuation, given a candidate signal amplitude and a noise power spectral density.

Finally,  we show two implications of this result for the interpretation of the double neutron star merger GW170817 \cite{2017LVCGW170817Discovery}. First, we estimate the size of current model uncertainties by evaluating the differences between 
waveform families in the posterior region of parameter space. We compare these to the envelope of calibration uncertainties and the differences within the posterior distribution of waveforms to assess the relative impact of waveform systematic error. Second, we use this method to interpret results from Edelman et. al. \cite{2021Edelman}, which constrain how the amplitude and phase of GWTC-1 signals depart from the waveform models used for analysis. Observed departures from the waveform models can be mapped to physical effects: unmodeled luminosity or internal energy contributions for the source system. This allows the largest median $\delta \phi$ at $\sim 60\,$Hz to be interpreted as an possible unmodeled energy transfer of $\delta E = \Delta E / E \sim 0.001$ relative to the orbital binding energy of $E \simeq -0.006 M_\odot c^2$.
In general, the recovered bounds on $\delta A$ and $\delta \phi$ limit the amount of unmodeled energy transfer compatible with the observed signals.

\section{Background}
\label{sec:background}
To interpret the observations of gravitational-wave astronomy, interferometer strain data $\bm{d} = \bm{h} + \bm{n}$ is assumed to be generated by an astrophysical strain signal $\bm{h}$ and background noise fluctuations $\bm{n}$. The likelihood of a detector measurement $\bm{d}$, given a particular incident wave $\bm{h}$, derives from the likelihood function for $\bm{n} = \bm{d} -\bm{h}$. In practice, it will be computed from the Fourier transforms of the data and signal time series, evaluated a set of discrete frequencies:
\begin{equation}
    p(\bm{d}|\bm{h}) \propto
    \exp\left(- \sum_i 
      2 \Delta f \frac{|d_i - h_i|^2}{S_n(f_i)}\right)
\end{equation}
where $S_n(f)$ is the power spectral density of the noise \citep{2020RomeroShawEt, 2015VeitchEt}. 

This motivates the definition of a noise-weighted inner product \cite{1994CutlerFlanagan} in either discrete or continuous form,
\begin{equation}
    \left<\bm{h_1},\bm{h_2}\right> = 2 \Delta f \sum_i \frac{h_{1i}^*h_{2i} + h_{1i} h_{2i}^*}{S_n(f_i)} = 4 \Re \int_0^{f_\text{max}} df \frac{\tilde{h}_1(f)^* \tilde{h}_2(f)}{S_n(f)} 
\end{equation}
This inner product definition sets $\left< \bm{n} , \bm{n}\right> = 1$. 
The expected signal-to-noise ratio $\varrho$ of an incident astrophysical signal $\bm{h}$ is given by
$ \varrho^2 = \left< \bm{h} , \bm{h} \right>$. We can then re-write the likelihood of the data $\bm{d}$ given an incident wave $\bm{h}$ as
\begin{equation}
    p(\bm{d}|\bm{h}) \propto \exp \left(-  \left<\bm{d} - \bm{h} , \bm{d} - \bm{h} \right>\right)
\end{equation}
Two waveforms are considered ``indistinguishable'' in a given detector if the difference between them is smaller than the noise: $\delta \bm{h} = \bm{h}_1 - \bm{h}_2$ satisfies $\left< \delta \bm{h} , \delta \bm{h} \right> < 1$. Often when comparing waveform families, especially those which agree at leading post-Newtonian order, we will have $\varrho_1 \simeq \varrho_2:=\varrho$, and indistinguishability is estimated by the ``mismatch'' after waveform differences are minimized over relative shifts in time and phase:
\begin{equation}
{\rm min}_{\Delta t_c, \Delta \phi_c}\left[ \left< \delta \bm{h} , \delta \bm{h} \right>  \right] \gtrsim 2 \varrho^2   \left( 1- {\rm max}_{\Delta t_c, \Delta \phi_c} \left[\left<\bf{h_1},\bf{h_2}\right>\right]
/\sqrt{\left<\bf{h_1},\bf{h_1}\right>\left<\bf{h_2},\bf{h_2}\right>}\right),
\end{equation} 
where the mismatch is in brackets on the RHS.
Gravitational waveform uncertainty requirements have thus been presented in terms of the mismatch between the true waveform and the model used when inferring source properties at a given signal-to-noise ratio $\varrho$ \citep{2008LindblomOwenBrown,2020PurrerHaster}. Mismatch is an integrated quantity over all frequencies.

In contrast, calibration uncertainties that affect the inferred $\bm{h}$ are explicitly computed as functions of frequency \cite{FarrTechnical,2020PayneTalbotThraneKissel,2021VitaleHasterSunFarrEt} and can be marginalized over for gravitational-wave inference \cite{2012VitaleDelPozzoLi,2020RomeroShawEt,2022EssickCalibration}. To do this, calibration uncertainty is cast in terms of amplitude and phase errors in the inferred detector strain compared to the true conditions of the local spacetime
\begin{equation}
\tilde{h}_\text{meas} = \tilde{h}_\text{true}(f)(1+\delta A_{\text{cal}}(f))\exp(i \delta \phi_{\text{cal}}(f))
\end{equation} 
which differentiate the measured $\tilde{h}_\text{meas}$ from the true strain $\tilde{h}_\text{true}(f)$ generated by an incident astrophysical wave. These corrections arise from the error budget of mapping between astrophysical and instrumentally measured strain \cite{2020SunEt,VirgoCalibration}. 

An approach similar to the calibration framework has been developed to constrain differences between the true signal and the waveform model used for inference: Edelman et al \cite{2021Edelman} estimated the size of unmodeled signal contributions  for the observations of GWTC-1 \cite{GWTC1}. To do this, coherent deviations across all detectors are also expressed in the form
\begin{equation}
    \tilde{h}_\text{astr}(f) = \tilde{h}_\text{model}(f)(1+\delta A(f))\exp(i \delta \phi(f))  
\end{equation}
where we are again using $\phi$ for the frequency-domain phase. In that work, corrections were modeled as splines for $\delta A(f)$ and $\delta \phi(f)$, and frequency-dependent departures from the baseline $\bm{h}$ model were constrained for signals observed by the LIGO \cite{Aasi_2015} and Virgo \cite{VIRGO:2014yos} observatories .

In this work, we demonstrate the frequency-dependent uncertainty in amplitude and phase $\delta A(f)$ and $\delta \phi(f)$ coming from existing waveform models. 
These uncertainties are explicitly connected to the underlying time-domain model uncertainties and the implications for the underlying physics of the source system. To do this, we follow standard procedures in the literature, but avoid common choices that assume a post-Newtonian framework with infinite coalescence frequency. We define all source characteristics in terms of weak-field gravitational-wave observables. Expressing model uncertainty in this form will allow explicit marginalization during the inference of source properties, and also allows the interpretation of waveform differences in terms of far-field source energetics.

\section{Waveform assumptions and model implications}
\label{sec:waves}

Waveform models come from time-domain physics of the emitting source and the response of the observatory's interferometer. Here, we review the mapping from the source properties to the signal prediction, to demonstrate how error intrinsic to the source physics can be disentangled from the reference time and phase $t_c$ and $\phi_c$ of a specific signal. 
The fundamental differences coming from varying model families are those that are independent of overall shifts in signal time and phase, and lead to a residual waveform model error that can be additionally marginalized over in inference.

We adopt a description of the time-domain waveform, similar to that in \cite{2022MezzasomaYunes}, with the assumption that we can write the source emission as a multipolar expansion of oscillatory mode functions.
Specifically, we assume the $h_+$ and $h_\times$ emitted along a line of observation from a given source can be characterized by an expansion in spin-weighted spherical harmonics \cite{2008Kidder, 2022GonzalezZappaBreschiBernuzzi}:
\begin{equation}
    h_+(t) - i h_\times (t) = 
        \sum_{\ell=2}^{\infty}
        \sum_{m=-\ell}^{\ell}
        h_{\ell m}(t) Y_{-2}^{\ell m}(\iota, \varphi)
\end{equation}
where $\iota$ is the inclination angle and $\varphi$ is the azimuthal angle of the line of sight from the source. Each $h_{\ell m}$ component is decomposed into a real time-domain amplitude and phase as
\begin{equation}
    h_{\ell m}(t) = \mathcal{A}_{\ell m}(t) e^{ - i \psi_{\ell m}(t)}
\end{equation}
in the usual form for characterizing gravitational radiation in numerical simulations,  described for example for the \texttt{lalsimulation} numerical relativity injection infrastructure \cite{2017SchmidtHarryPfeiffer} following numerical data formats of \cite{2007AjithBoyleBrownEt}. We assume the time-domain amplitude scales as $\mathcal{A} (t)= \mathcal{A}_{0}(t) \left(d_0/d\right)$ for a source at luminosity distance $d$ in terms of $\mathcal{A}_{0}(t)$ at a reference distance $d_0$ \cite{2021ChenHolzEvansVitaleCreighton}.

The astrophysical strain $\bf{h}$ measured by a single detector is a projection on to the detector frame of the incident time domain polarizations $h_+(t)$ and $h_\times(t)$,
\begin{equation}
    h(t) = F_{+}(\alpha, \delta, \psi_{p})h_+(t)
    + F_{\times}(\alpha, \delta, \psi_{p})h_\times(t)
\end{equation}
with the specific detector's antenna response functions $F_{+,\times}$ that depend on the source's sky location (right ascension $\alpha$ and declination $\delta$) and a polarization orientation angle $\psi_p$ of the source relative to the interferometer. In a single detector observation, the polarization angle is degenerate with phase. A sky location directly above or below an interferometer with orthogonal arms gives $F_+^2 + F_\times^2=1$.


The incident $\bm{h}$ can  be written as the real part of a sum over the spherical harmonic mode amplitudes. 
\begin{equation}
    h(t) = \sum_{\ell m} Q_{\ell m} h_{\ell m}
\end{equation}
where $Q_{\ell m}$ captures the sky-location-dependent detector response to each mode given the source orientation. $Q_{\ell m}$ will be constant for short-duration transients; for long-duration signals a transfer function encoding the response of the detector would need to be included in the signal model \cite{2018MarsatBakerFDMod}.



The leading-order quadrupole modes $\ell,m = 2,\pm2$ are dominant for many gravitational-wave sources. We consider circular, non-precessing, near-equal-mass binaries as the sources in this demonstration, so will restrict to this case --- neutron-star binary signals have masses within a range of roughly 1--2.5$\,M_\odot$ \cite{GWTC3Pop:2021} and are well-described by the $2,\pm2$ mode.
For an optimal sky location directly above or below the detector, with detector arms aligned with the plus polarization, the sum of the $\ell,m = 2,\pm2$ modes gives 
\begin{equation}
h_+(t) - i h_\times(t) = \sqrt{5/16\pi}\left(\left(1+\cos^2 \iota\right) \cos 2 \varphi + 2 i \cos \iota \sin 2 \varphi \right) h(t)
\end{equation}
which yields $h(t)$ from $h_{22}(t)$ with the conversion factor $Q_{22} = \sqrt{5/4\pi}$ for an optimally oriented, face-on source. If a general sky location and inclination are considered,
$Q_{22} = \sqrt{5/4\pi}\left(F_+^2 \left( 1+ \cos^2 \iota \right)^2 /4 + F_\times \cos^2 \iota \right)^{1/2}$. Overall, measured signal amplitude for quadrupole sources will be scaled relative to the optimal face-on and overhead configuration by the effective distance $d_\text{eff} = d \left(F_+^2 \left( 1+ \cos^2 \iota \right)^2 /4 + F_\times \cos^2 \iota \right)^{- 1/2}$ which combines the effects of luminosity distance, sky location, and the source's orientation angles \cite{2012AllenAndersonBradyBrownCreighton}. 

We focus now on signals from inspiraling compact binaries, where the frequency of the signal sweeps slowly upward as the binary evolves over multiple cycles. 
We are interested in modeling the uncertainty given known physical properties of the source --- for example, those characterizing the masses, spins, eccentricity, and tides --- which we call intrinsic properties. These properties will determine the amplitude of gravitational waves as function of the emission frequency, and the gravitational-wave emission also drives a change in the emission frequency. We show how to compute the Fourier-domain signal $\tilde{h}(f)$, with $f$ the Fourier frequency, from integrations of the instantaneous gravitational-wave frequency $F = \dot\psi/ 2 \pi$ of the time-domain phase. This results in the integration constants $t_c$ and $\phi_c$ that fix the arrival time and phase of a specific source. The intrinsic properties of the source model generate the characteristic $\dot F (F)$.

A relatively slowly-varying amplitude
allows the use of the stationary phase approximation (SPA) in determining the frequency-domain Fourier transform for each $m>0$ and $f>0$ \cite{1994CutlerFlanagan,1999DrozKnappPoissonOwen,2022MezzasomaYunes} 
when the two conditions
\begin{align}
\label{eq:spa-conditions}
    \left| \frac{d}{dt} \ln \mathcal{A}_{\ell m}(t) \right|
    &\ll \left| \dot{\psi}(t) \right|  &|\ddot \psi| &\ll \dot \psi^2
\end{align}
are satisfied. Overdots denote a time derivative. 
Error from the SPA is expected to be smaller than windowing artifacts for gravitational-wave inspirals \cite{1999DrozKnappPoissonOwen}. For the reference neutron-star binary system used later in this work, all waveforms including those from numerical simulation will satisfy LHS$< 0.1\times$RHS for both conditions, only reaching 0.1 as the stars collide. This supports the use of instantaneous frequency as a characteristic of the time-domain system. The SPA has been shown to give faithful frequency-domain representations for state-of-the-art time-domain waveforms up to merger \cite{GambaBernuzziNagar2021FFT}.  Applications to LISA signals  demonstrate higher order contributions that could be used for extension through the merger phase \cite{Hughes:2021exa}.

To apply the SPA, we expand the transform integral around time where the instantaneous frequency $F$ of the incident wave matches the Fourier frequency $f$ of interest, specifically  $T_{\ell m}$ defined by
\begin{equation}
    \dot \psi_{lm} \left(T_{\ell m} \right) =  2 \pi f 
\end{equation}
The  resulting frequency domain waveform is $\tilde{h}(f) = A(f) \exp (- i \phi(f))$, with
\begin{subequations}
\label{eq:spa}
\begin{align}
    A(f) &= 
    \sum_{\ell m}
    Q_{\ell m}
    \left(
    \frac{2\pi} { \ddot\psi_{\ell m}(T_{\ell m})} 
    \right)^{1/2}
    \mathcal{A}_{\ell m}(T_{\ell m})
   \\
   \phi(f) &= \frac{\pi}{4} +
    \psi_{lm}(T_{\ell m})  
   - 2 \pi f T_{\ell m}
\end{align}
\end{subequations}
The SPA allows direct use of model predictions for the time-domain amplitude and phase $\mathcal{A}_{\ell m}(t)$ and $\psi_{\ell m}(t)$ to calculate the  corresponding frequency-domain waveform.

The functions that enter into $\tilde{h}$ are relative to an explicit coalescence time $t_c$, which  is traditionally defined in the $f\to\infty$ limit following Cutler and Flanagan \cite{1994CutlerFlanagan}. Since $f\to\infty$ does not happen in all models, we here define this as the time at which the chosen waveform model reaches a specific reference coalescence frequency $f_c$.  The LIGO/Virgo software \texttt{lalsimulation} instead uses a convention where $t_c=t_\text{peak}$ is defined by the maximum amplitude of the waveform \cite{2017SchmidtHarryPfeiffer}. This choice will imply a reference $f_c = F(t_\text{peak})$ which varies between waveform models. For fixed $t_c$ and $\phi_c$, differences between waveform models will depend on the $f_c$ assumed in the parameter estimation, so will be sensitive to the definition of $t_c$. However, the residual phase error as defined below will be independent of $f_c$.

For each time-domain waveform model, the following  time and phase observables are generated from the wave's instantaneous frequency $F$ and its time derivative $\dot F$,
\begin{subequations}
\label{eq:Tpsidef}
\begin{align}
    T(f) &= t_c - \int_{f}^{f_c} dF\, T'(F)\\
    \psi(f) &= \psi_c 
                - 2 \pi \int_{f}^{f_c} dF\, F\, T'(F) 
           \\
            &= \psi_c - 2 \pi \left(  f_c t_c  - f T(f)
            - \int_{f}^{f_c} T(F) dF \right)
\end{align}
\end{subequations}
defining the function $T'(F) := \left(\dot F(F) \right)^{-1} = d\, T(F) /d F$ to streamline notation. These observables are entirely defined from the far-field wave emission, and can be interpreted as $t_c - $(time to $f_c$) and $\psi_c - $(time-domain phase accumulation remaining before $f_c$). After the instantaneous frequency is calculated, $(T-t_c)$ and $(\psi-\psi_c)$ can be read in directly from numerical simulation data.

Substituting $T(f)$ and $\psi(f)$ into Eqs.~\ref{eq:spa} yields the frequency-domain form for the quadrupole waveform in terms of the characteristic functions, as we begin to write $\mathcal{A}(f)$ for $\mathcal{A}(T(f))$:
\begin{subequations}
\begin{align}
    A(f) &= 
    Q(\bm{\theta}_\text{ext}) 
    \left(
   T'(f)
    \right)^{1/2}
    \mathcal{A}(f) /2\\
   \phi(f)  
     &= \frac{\pi}{4} + \psi(f) - 2\pi f T(f)  \\  
    &= \frac{\pi}{4} + \psi_c - 2 \pi \left( f_c t_c - \int_f^{f_c} dF \,T(F) \right) \label{eq:strainphic}\\
    &= \phi_c - 2\pi f t_c + 2\pi \int_f^{f_c} d \tilde{f} \int_{\tilde{f}}^{f_c} dF \, T'(F)  \label{eq:phasemarg}\
\end{align}
\end{subequations} 
Setting $f=f_c$ in Eq.~\ref{eq:strainphic} defines the signal coalescence phase $\phi_c = \frac{\pi}{4} + \psi_c - 2 \pi f_c t_c$. Note that $\phi_c=0$ does not correspond to $\psi_c=0$. The explicit second integration over frequency $\tilde{f}$ demonstrates  how $\phi(f)$ depends on the underlying model's $T'(F)$ and the two integration constants $t_c$ and $\phi_c$ in Eq.~\ref{eq:strainphic}.  The $t_c$ and $\phi_c$ that characterize an observed signal emerge as integration constants in Eq.~\ref{eq:phasemarg}, a form that allows marginalization over reference phase $\phi_c$ and time $t_c$ in gravitational-wave inference \cite{VeitchPozzo,2020RomeroShawEt}. The remaining intrinsic parameter contribution is entirely from $T'(F)$.

Following the form used in calibration marginalization, we write frequency-domain signal uncertainty terms as $\delta A(f)$ and $\delta \phi(f)$
\begin{equation}
    \tilde{h}_\text{true}(f) = \tilde{h}_\text{model}(f)(1+\delta A(f))\exp(i \delta \phi(f))
\end{equation}

The intrinsic model dependence of $\tilde{h}$ will be derived from the characteristic $\mathcal{A}$ and $T' = 1/\dot{F}$ functions of the waveform family in question. We can therefore express the impact of modeling differences on the signal in terms of these functions, specifically in the form
\begin{subequations}
\label{eq:modelerror}
\begin{align}
\mathcal{A}_\text{true} (f) &= \mathcal{A}_\text{model} (f) \left(1+\delta \mathcal{A} (f)\right),\\
T'_\text{true} (f) &= T'_\text{model} (f) (1 +  \delta T' (f)) 
\end{align}
\end{subequations}

It is computationally useful to track differences between models through the intermediate functions $T(f) = T_\text{model} (f) + \delta T (f)$ and $\psi(f) = \psi_\text{model}(f) + \delta \psi(f)$.
As the contribution to $T(f)$ and $\psi(f)$ are defined relative to $f_c$ in Eqs.~\ref{eq:Tpsidef}, the differences $\delta T(f)$ and $\delta \psi(f)$ will also be relative to $f_c$. The signal uncertainty at fixed $t_c$ and $\phi_c$ is 
\begin{subequations}
\label{eq:signal-from-model-uncertainty}
 \begin{align}
     1+ \delta A (f) &= \left(1+ \delta T'(f)\right)^{1/2} \left(1 + \delta \mathcal{A}(f)\right)\\
     \delta \phi (f) 
     &=  \delta \psi(f) - 2 \pi f \delta T(f) 
   =2 \pi \int_f^{f_c} d \tilde{f} \int_{\tilde{f}}^{f_c} dF \, T'(F) \delta T'(F)
\end{align}
\end{subequations}
 
The historic choice of $f_c \to \infty$ results in better-behaved modeling uncertainties at fixed $t_c$ and $\phi_c$ for post-Newtonian expansions than lower reference frequencies.

To determine the impact of model uncertainties on inference, it's important to consider that the integration constants $\phi_c$ and $t_c$ will be searched over when identifying incident waves and marginalized over when determining the intrinsic source properties \cite{2020RomeroShawEt, FarrTechnical,VeitchPozzo}. The impact of waveform uncertainty on measurements will be reduced by this marginalization.
For example, a toy modified-GR waveform model that produced only an overall constant shift $\delta \phi(f) = \delta \phi_0$ across all frequencies would not change any recovered parameters other than $\phi_c$. Similarly, an overall shift of the form $\delta \phi(f) = 2 \pi f t_0$ would be entirely absorbed by marginalization over $t_c$. 

To find the residual phase error for a measurement scenario, consider a model $\tilde{h}(f)$ with errors $\delta A(f)$ and $\delta \phi(f)$. The amplitude $\delta A(f)$ in the Fourier domain is independent of shifts in time and phase. However, when inferring the properties of this model from a signal that differs by $\delta \phi(f)$, any model uncertainty described by $\phi_0 + 2 \pi f t_0 $ will be absorbed by marginalization over shifts in time and phase during inference. In the absence of noise, the maximum likelihood value of $\phi_0$ and $t_0$ will be at the minimum of 
\begin{subequations}
\begin{align}
\left< h - h_\text{model} | h - h_\text{model}  \right> &\propto
  \sum_i \frac{
    \left| A(f_i)\right|^2 \left| 1-
    \exp{i \left( \delta\phi(f_i) 
    - \phi_0 - 2\pi f_i t_0\right) }\right|^2
    }{S_n(f_i)} \\
    &\simeq \sum_i \frac{
    \left| A(f_i)\right|^2}{S_n(f_i)} \left|  \delta\phi(f_i) 
    - \phi_0 - 2\pi f_i t_0 \right|^2
\end{align}
\end{subequations}
where the approximation applies when residual model error is small enough that that a small angle approximation can apply across relevant frequencies.
 The maximum likelihood condition becomes equivalent to a weighted least squares fit of $\delta \phi (f)$ to the linear function $\phi_0 + 2 \pi f t_0 $, with weights $|A(f)|^2 / S_n(f)$ corresponding to the expected variance at each frequency. Subtracting off the fit leaves a residual model error $\delta \phi_\text{res}$. This characterizes the frequency-dependent waveform phase uncertainty that should be marginalized over in addition the existing time and phase marginalization.

\section{Uncertainty estimates for current models}
\label{sec:currentmodels}
A proper error estimate should rise from careful analysis of the range of viable analytic choices, universal relation uncertainties, numerical resolution error, neglected effects in modeling, or other choices made by the waveform modelers. The effect of a particular range of ``reasonable'' choices will determine the frequency-dependent error distribution of the model-characteristic functions $T'(F)= 1 / \dot F (F)$ and $\mathcal{A}(F)$, as in Eqs.~\ref{eq:modelerror}, where $F$ is defined as instantaneous frequency from the phase evolution of the waveform model. These distributions will best be estimated directly by those creating state-of-the-art models, and could be released in concert with a waveform implementation.  

To illustrate the size of current modeling uncertainties, however, we show an example comparing  waveform models from \texttt{lalsimulation} \cite{lalsuite} and simulations from the \texttt{CoRe} library \cite{2022GonzalezZappaBreschiBernuzzi}. This illustrates both the current model differences as well as a possible application of this error budget calculation $A(f)$ and $\phi(f)$. 

We start with a high-resolution \texttt{CoreDB} numerical simulation waveform from Dietrich et. al., \texttt{BAM:0095} \cite{Dietrich:2017aum}, which has masses $m_1=m_2=1.349998$, zero spin, and the SLy equation of state. This equation of state has $\Lambda_1=\Lambda_2=390.1104$ at this mass, and we
 generate tidal waveform models with matching $\bm{\theta}_\text{int}$ parameters using \texttt{TEOBResumS} \cite{Nagar:2018plt,Akcay:2018yyh} and \texttt{SEOBNRv4T} \cite{2021SteinhoffHindererDietrichFoucart,2016HindererTaracchiniFoucart}.
We calculate instantaneous frequency with second order accurate central differences using \pyth{numpy.gradient} for each model. For the numerical simulation data, we fit $f(t)$ with a B-spline using \pyth{scipy.interpolate} with smoothing to reduce sub-orbital oscillations in the derivative quantity. Another \pyth{gradient} gives $\dot{F}$. 

We show also the equivalent model characteristics of the \texttt{IMRPhenomP} model \cite{HannamSchmidtBohe2014} with numerical tidal contributions \texttt{NRTidal} \cite{DietrichKhanDudiEt2019,DietrichSamajarEt2019} as well as the \texttt{TaylorF2} \cite{BuonannoIyerEt2009} waveforms with post-Newtonian tides \cite{VinesFlanaganHinderer} using the analytic stationary phase approximation inversion
\begin{subequations}
\label{eq:freqdomaintotimedomain}
\begin{align}
    T(f) &= \phi' (f) / (2 \pi) - t_c^\prime \\
    \psi(f)  &= \phi(f) + 2 \pi f \, T(f) - \phi_c^\prime \\
    \dot F(F) &= \frac{1}{T'(F)} = \frac{2 \pi}{\phi''(F)}\\
    \mathcal{A}(F) &= \frac{1}{Q} 
    \sqrt{\dot F (F)}
    \left|\tilde h(F) \right|
\end{align}
\end{subequations}
where we choose $\phi_c^\prime$ and $t_c^\prime$ to set the $T$ and $\psi$ functions to zero at our chosen $f_c$.

\begin{figure}[htb]
    \centering
    \includegraphics[height=2.2in]{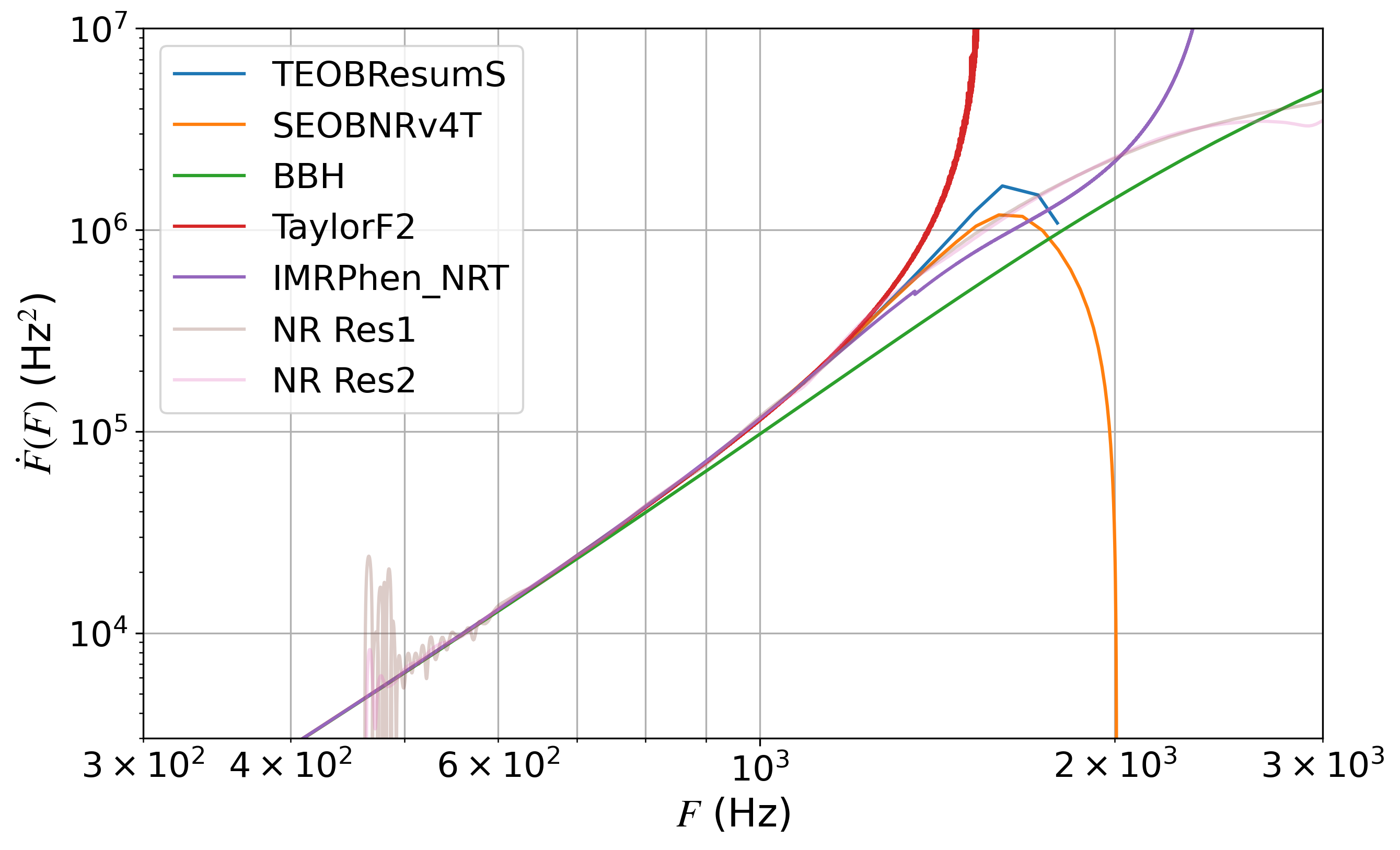}
    \includegraphics[height=2.2in]{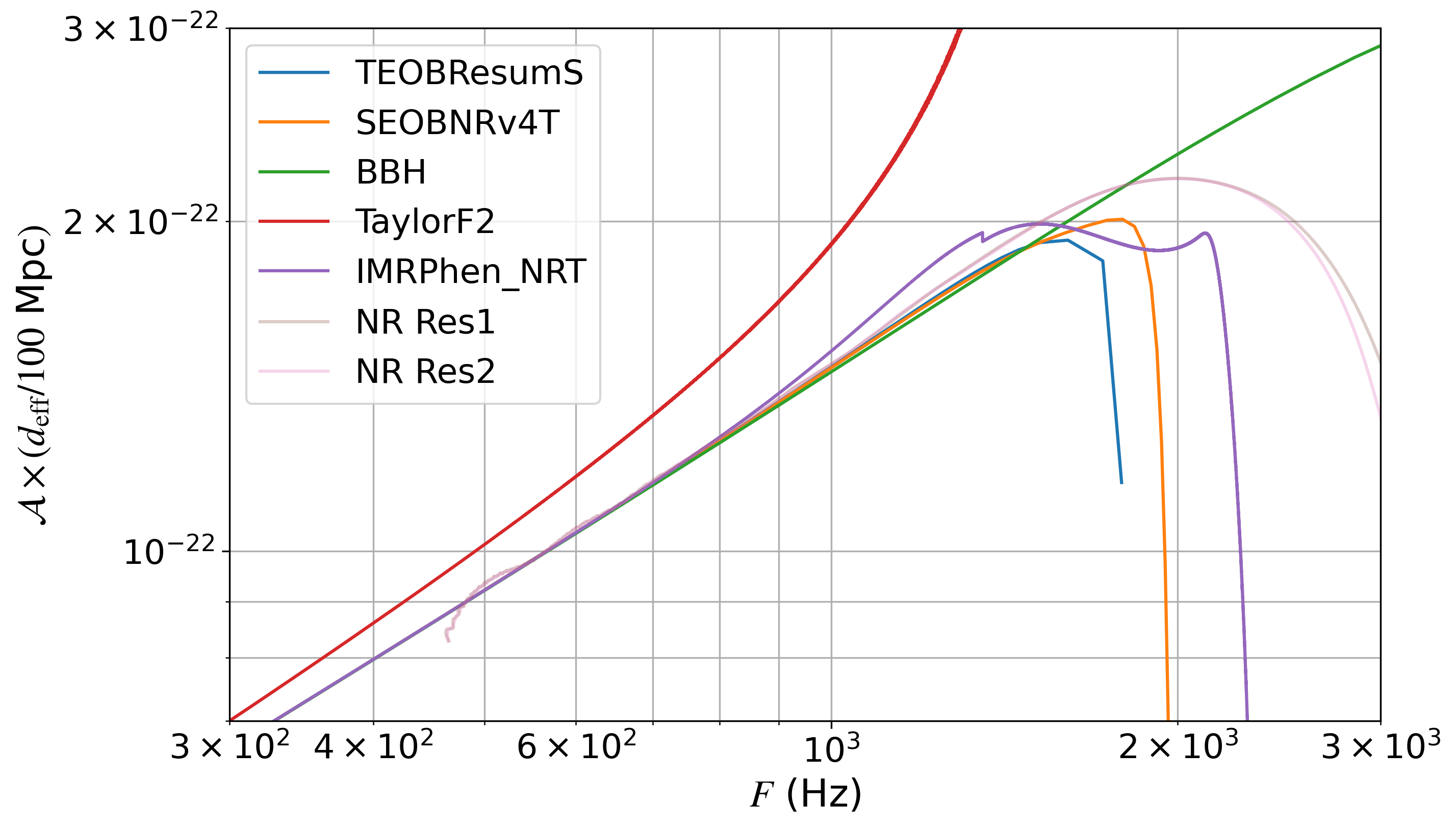}
    \caption{The characteristic functions $\dot F (F)$ and the time-domain strain amplitude $\mathcal{A} (F)$ as a function of instantaneous frequency.   These functions characterize the waveform predictions of each model independent of overall shifts in time and phase. Results are shown for a set of time-domain models, frequency-domain models, and two resolutions of numerical simulation data. Some nonphysical oscillation is visible for the numerical data from the start of the simulations. All simulate a zero-spin double neutron star system with $m_1=m_2=1.35 M_\odot$ and the SLy equation of state, except for the binary black hole (BBH) which uses \texttt{SEOBNRv4} \cite{Bohe:2016gbl} for the same masses. The difference in time-domain amplitude $\mathcal{A}$ for \texttt{TaylorF2} at these frequencies is because it includes only the leading-order amplitude term. The two light-shaded NR simulation resolutions demonstrate consistent frequency-domain amplitude and phase predictions from early tidal departure through to merger frequencies. 
    }
    \label{fig:FFdotf}
\end{figure}

Fig.~\ref{fig:FFdotf} shows the resulting $\dot F$ and $\mathcal{A}$ for all models. As will be discussed in Section \ref{sec:energetics}, differences in $\dot F$ can be interpreted in terms of additional energy losses as a function of frequency, for example those that rise from tidal contributions relative to the BBH model. The amplitude $\mathcal{A}$ is connected to the gravitational-wave luminosity at a give frequency, with differences arising for example from changes in the source quadrupole moment.

We note that this comparison of numerical data and semi-analytic models is independent of any waveform alignment choices such as shifts in time and phase; it is closely related to comparisons of orbital frequency derivative used in numerical simulation analyses (e.g. \cite{2012BernuzziNagar}) cast in terms of waveform characteristics. Fitting these functions directly from a combination of semi-analytic and numerical-simulation information could therefore inform a numerical-relativity calibrated waveform model which is independent of hybridization choices and extends through to the post-merger phase.


For this set of models, we note that \texttt{TEOBResumS} terminates at its $t_c$ with a lower $f_c$ than other approximants. For the first phase comparison, we therefore choose \texttt{TEOBResumS} as our baseline and set $f_c = f_\text{peak,TEOB} \simeq 1787$\,Hz as a reference frequency which can be chosen to be common for all models.

\begin{figure}[htb!]
\begin{center}
\includegraphics[width=0.49\textwidth]{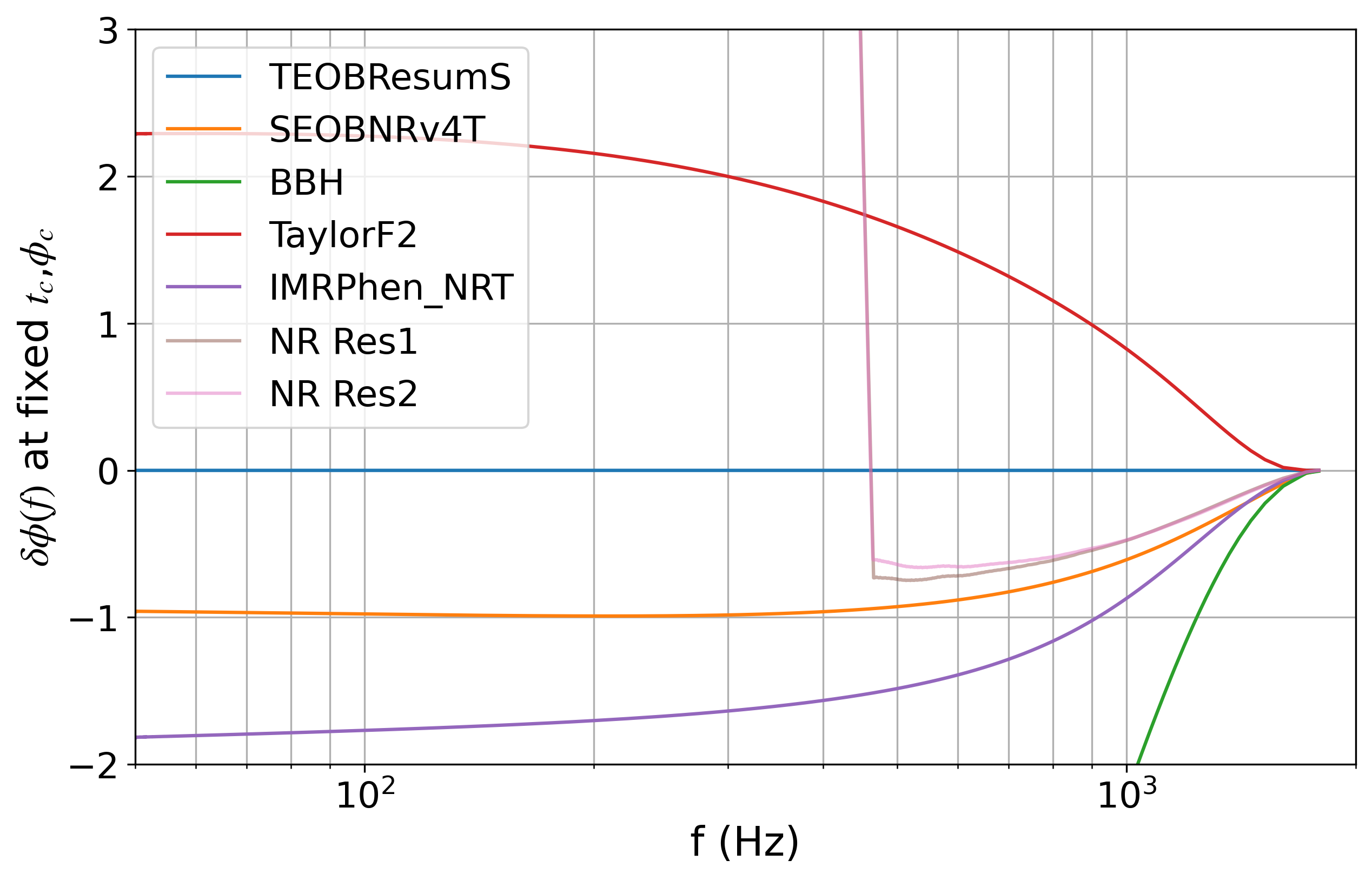}
\end{center}
\caption{Phase differences of the predicted waveform $\tilde{h}$, relative to the \texttt{TEOBResumS} model.  Phase is measured relative to $f_c = 1783$ as defined for \texttt{TEOBResumS}, so all models have $T(f_c)=0$ and $\phi(f_c)=0$ at this frequency. All simulations and models have the same mass and tide parameters, except for BBH which sets the tides to 0.}
\label{fig:signal-error}
\end{figure}

We compute $T(f)$ and $\psi(f)$ relative to the common reference frequency and regenerate the Fourier-domain phase.  Although smoothing was require to reduce differentiation noise seen in Fig.~\ref{fig:FFdotf}, the integrated $\psi(f)$ relative to the reference frequency can be read directly from numerical data. This gives from each waveform the amplitude and phase of a Fourier-domain signal $\tilde{h} (f)$ with a consistent $t_c$ and $\phi_c$, using the stationary phase approximation of Sec.~\ref{sec:waves}. Once the signal-domain functions are defined, we can compute relative amplitude error  $\delta A$ and the phase shift $\delta \psi$ relative to the baseline \texttt{TEOBResumS} model that set $f_c$. The phase error resulting from waveform differences is relative to the reference frequency, and naturally goes to zero at that frequency.  Reults are shown in Fig. \ref{fig:signal-error}. 

\begin{figure}
    \centering
    \includegraphics[width=0.49\textwidth]{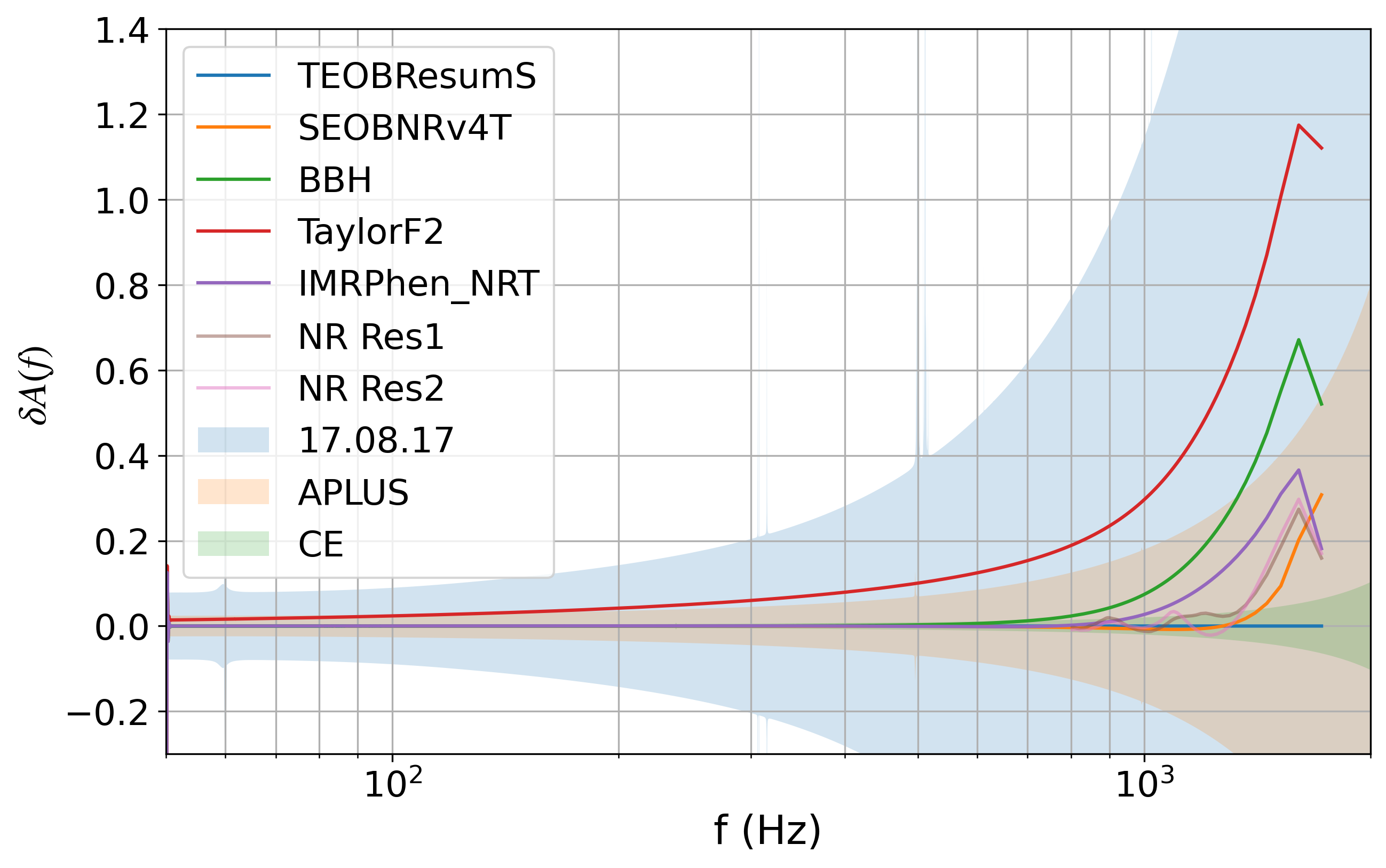}
    \includegraphics[width=0.49\textwidth]{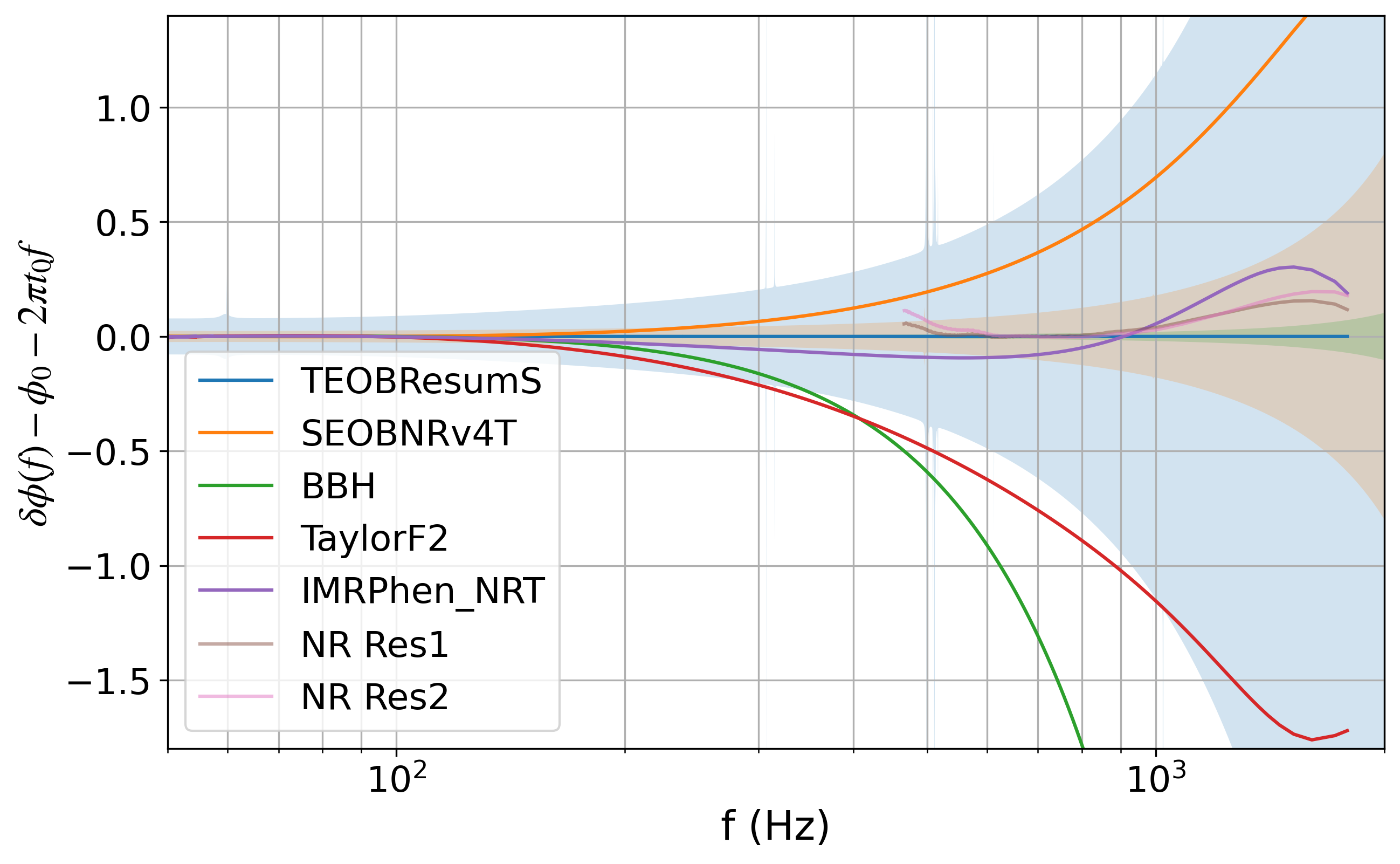}
    \caption{Relative signal amplitude differences $\delta A(f)$  and residual signal phase differences $\delta \phi_\text{res}(f)$ of waveform models relative to the reference model \texttt{TEOBResumS}. Residual phase is that remaining after removing a least-squares fit of $\delta \phi (f) = \phi_0 + 2 \pi f t_0 $ weighted by the expected variance $\propto A(f)^2/S_n(f)$ in A+; the choice of ground-based detector has minimal impact on the result. Numerical simulations are too short to compute residual phase; phase differences for both resolutions from \texttt{TEOBResumS} with $f_c=700$\,Hz are shown for reference. The estimated equivalent contribution of noise fluctuations in current and next-generation detectors for a signal at
    $d_\text{eff}=100$\,Mpc is shown following discussion of Section \ref{sec:requirements}.}
    \label{fig:dph-compare-detector}
\end{figure}

To determine measurement impact here, we calculate the residual phase differences that must remain after marginalization over time and phase. We calculate residual phase differences by removing a weighted least squares fit of $\delta \phi(f)$ to $\phi_0 + 2 \pi f t_0$ with variance $S_n(f)/A^2(f)$, and show the resulting differences with the ``APLUS'' $S_n$ \cite{T1800042-APLUS} in Fig.~\ref{fig:dph-compare-detector}. 
The residual phase error is not very sensitive to the specific ground-based detector noise spectrum used to fit $\phi_0$ and $t_0$; very similar residual phase is found using $S_n(f)$ from other ground-based spectra. 
As a cross-check, recalling that most of the signal-to-noise accumulates through the early inspiral \cite{2021DietrichHindererSamajdar} and would ``anchor'' the $t_c$ and $\phi_c$ used to describe the models, we verified that similar residual error can also be found by subtracting a linear fit to $\delta \phi$ in the ``bucket'' of ground-based detector sensitivities, between 50 and 150 Hz.

Calibration envelopes from O4 \cite{2020SunEt} give $\delta A$ better than 11.29\% and $\delta \phi$ better than 9.18 deg or 0.160 radians from 20-2000 Hz. Comparing the results of Fig. \ref{fig:dph-compare-detector}, we see that waveform uncertainties for these neutron-star models will be smaller than calibration uncertainties at low frequency. Current waveform differences for $\delta \phi$  between \texttt{TEOBResumS} and \texttt{SEOBNRv4} may exceed calibration error above 500 Hz and between \texttt{IMRPhenomD\_NRTidalv2} and \texttt{TEOBResumS} above 1 kHz. Waveform differences from $\delta A$ remain below calibration uncertainty past 1 kHz for tidal models that go beyond leading order.

\section{Detector-dependent accuracy requirements for calibration and modeling}
\label{sec:requirements}
For our reference $\delta A(f)$ and $\delta \phi(f)$, we can compute the resulting waveform differences and estimate whether they have the potential to affect measurements in a detector with a given noise power spectral density $S_n(f)$ and signal amplitude $A(f)$.
This motivates frequency-dependent modeling requirements to bring  systematic uncertainties below the level of frequency-dependent statistical fluctuations implied by a future detector's power spectral density $S_n(f)$ for a given signal. These criteria will apply equally well to the $\delta A(f)$ and $\delta \phi(f)$ of calibration envelopes, similarly applied to the amplitude of a specific signal.

We choose three reference noise PSDs in this work. First, we denote as ``17.08.17'' the LIGO Livingston power spectral density during the observation of GW170817 \cite{2019LVKGW170817Properties}, as documented in \cite{P1800061-170817}, to set the scale of background noise in 2017, and compare with assessments of waveform error in determining GW170817's source properties.
Second, we show as ``APLUS'' the A+ detector design of  \cite{T1800042-APLUS}, as a reference for the goal sensitivities of LIGO/Virgo/Kagra's O5 observing run in the later 2020s \cite{2018LRRProspects,Aasi_2015,VIRGO:2014yos}. Finally, as a reference for what might be anticipated for next-generation gravitational-wave observatories taking data in the 2030s, we show results using a wide-band Cosmic Explorer noise configuration  ``CE'' \cite{P1600143-CE, 2019ReitzeCE,CE-Horizon}, which has comparable sensitivity to the Einstein Telescope \cite{Maggiore_2020}.

\begin{figure}
    \centering
    \includegraphics[width=0.49\textwidth]{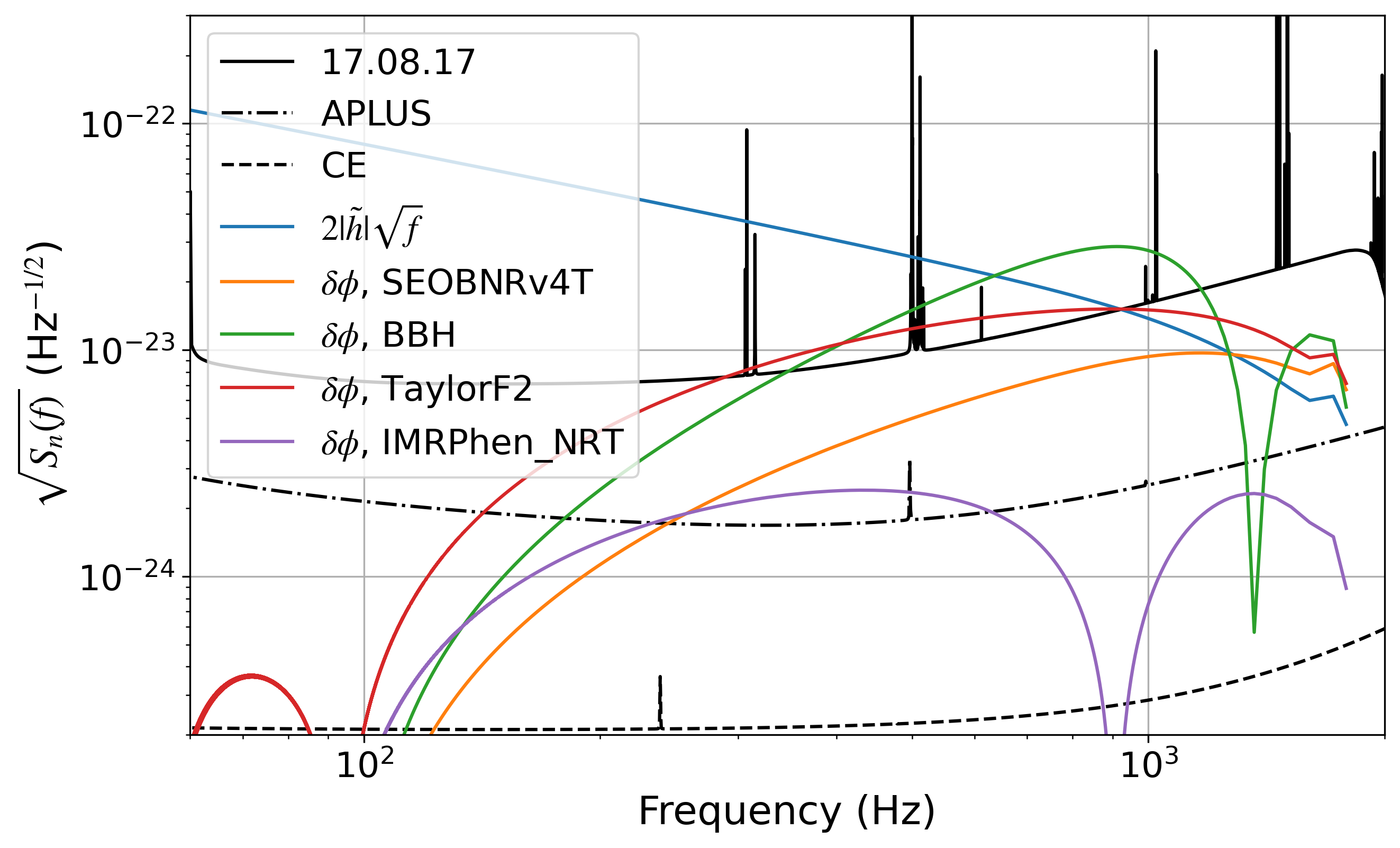}
    \includegraphics[width=0.49\textwidth]{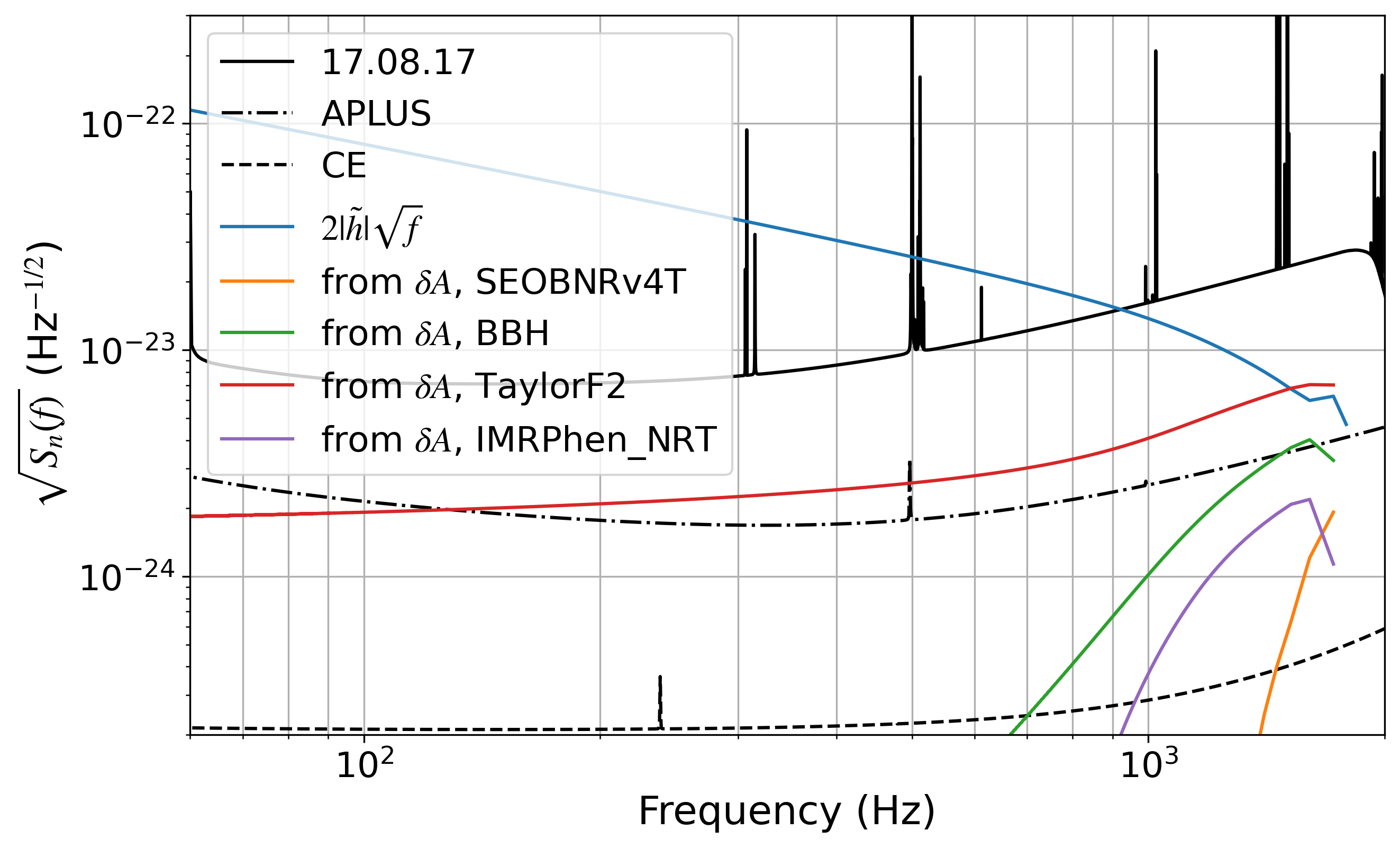}
    \caption{Waveform differences of semi-analytic waveform models from phase differences relative to \texttt{TEOBResumS} for a signal at $d_\text{eff} = 100\,$Mpc.
    This uses the same $\delta A(f)$ and $\delta \phi (f)$ as Fig.~\ref{fig:dph-compare-detector}, with its each effect propagated through to $|\delta h(f)|$, including for $\delta \phi(f)$ the motion in and out of phase in $ A(f) \sqrt{2(1- \cos \delta \phi(f))}$ which gives the bumpy structure. A similar plot style was used in \cite{2013ReadBaiottiCreighton} to compare waveform differences to detector noise. A ratio of the noise and signal quantities squared is integrated to determine whether waveforms are `distinguishable' by the criteria of Ref.~\cite{2008LindblomOwenBrown}.}
    \label{fig:strain-difference-compare-detector}
\end{figure}

Recall that the signal coming from the difference between two waveforms is $\delta \bm{h} = \bm{h_1} - \bm{h_2}$. 
In general, we are interested in the magnitude of the waveform difference $\sqrt{\left<\delta \bm{h}, \delta \bm{h}\right>}$, which is the signal-to-noise ratio of the waveform differences. For a relative difference in amplitude $\delta A (f) $, this is given by
\begin{equation}
\left<\delta \bm{h}, \delta \bm{h}\right>= 4 \int_0^{f_\text{max}}df\, \frac{|\delta \tilde{h}(f)|^2}{S_n(f)}= 4 \int_0^{f_\text{max}}df\, \frac{A^2(f) \delta A^2(f)}{S_n(f)}
\end{equation}
and for a frequency-dependent phase difference $\delta \phi (f) $ we have
\begin{subequations}
\begin{align}
\left<\delta \bm{h}, \delta \bm{h}\right> 
&= 4 \int_0^{f_\text{max}} df\, \frac{A^2(f) \left|1 - \exp{\left(i \delta \phi (f) \right)}\right|^2}{S_n(f)} \\
&= \int_0^{f_\text{max}} df\, \frac{2 A^2(f) \left(1 - \cos \delta \phi \right)}{S_n(f)} 
\end{align}
\end{subequations}

Requirements for $\delta\phi (f)$ and $\delta A(f)$ are set from a particular level of background noise fluctuations $S_n(f)$. 
The frequency-bin variance $f S_n(f)$ comes from the power spectral density $S_n(f)$.  The characteristic strain squared of the signal in the frequency bin is $4 f^2 |\delta \tilde{h}(f)|^2$, and is defined so that the signal-to-noise of the waveform difference $\left<\delta \bm{h} | \delta \bm{h}\right>$ will be less than one, and the waveforms are indistinguishable as defined in Section \ref{sec:background}, if the characteristic strain $4 f^2 |\delta \tilde{h}(f)|^2 < f S_n(f)$ at all frequencies - see \cite{2015MooreColeBerry,2004CampCornish} for explanations of characteristic strain.

We show this comparison recast in terms of the more commonly plotted amplitude spectral density $\sqrt{S_n(f)}$ and the effective strain amplitude 
$2 \sqrt{f} |\delta \tilde {h}|$ in Figure \ref{fig:strain-difference-compare-detector} for a \texttt{TEOBResumS} reference signal at $d_\text{eff}=100\,$Mpc. All waveform parameters are set following the reference numerical simulation of Section \ref{sec:currentmodels}.
While the \texttt{IMRPhenomPv4\_NRT} waveform model remains nearly indistinguishable from \texttt{TEOBResumS} even for A+ sensitivites, other waveform models have differences with a potential impact on the likelihood. However, as explored in \cite{2020PurrerHaster}, the actual impact on the inferred parameter posterior distributions depends additionally on whether or not the model differences (or calibration errors) are orthogonal to the waveform differences introduced by a parameter variation.
 The result is similar to the waveform difference estimates of \cite{2013ReadBaiottiCreighton}, but uses the maximum-likelihood phase and amplitude method of Section~\ref{sec:waves}. In this work, we can recover a similar estimate of systematic uncertainty for state-of-the-art analytic model variants that do not share a common $f_\text{peak}$.

\begin{figure}[tb]
    \centering
    \includegraphics[width=0.6\textwidth]{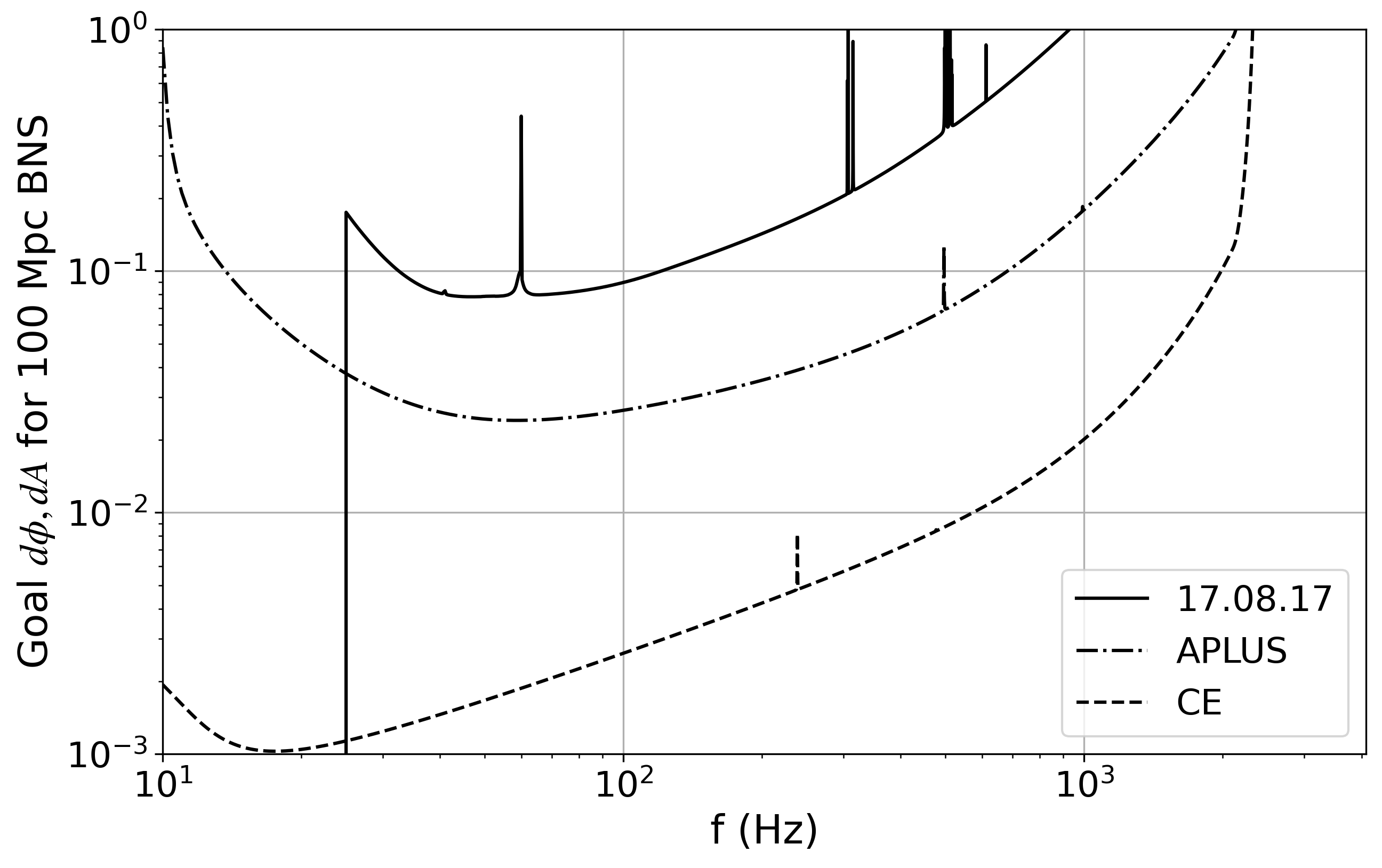}
    \caption{The size of of $\delta \phi$ (in radians) or $\delta A$ (as a fraction) that gives signal modifications with the same characteristic strain as that from the power spectral density for the labeled detectors, for an incident BNS signal from $d_\text{eff}=100$\,Mpc. The model uncertainty and calibration requirements set by the power spectral density are specific to the signal under consideration; the rapid rise at 2 kHz for CE reflects the termination frequency of the inspiral model being applied here. The important region for measurement and calibration will be set by the frequency range of the desired signal information, as could be determined for example in Figure 2 of Ref.~\cite{2021DietrichHindererSamajdar}. For signals of this amplitude, systematic uncertainties from calibration and waveform modeling seem likely to dominate the error budget at $\sim 10-100$ Hz in next-generation observatories.}
    \label{fig:goal-error}
\end{figure}

For $\delta A(f)$, and for sufficiently small $\delta \phi(f)$ that $\sqrt{2 (1-\cos \delta \phi)}\simeq \delta \phi$, this comparison motivates a goal for the allowable waveform and calibration errors by again rearranging the frequency-by-frequency indistinguishability criterion to
\begin{align} 
|\delta A(f)| &< \frac{\sqrt{ S_n(f)}}{2 A(f) \sqrt{f}}
& \text{and} & &
 |\delta \phi(f)| &< \frac{\sqrt{ S_n(f)}}{2 A(f) \sqrt{f}}
\end{align}
For both waveform and calibration uncertainties, then we have a goal frequency-
dependent tolerance that is set by the noise spectral density and waveform amplitude. 
We  show the frequency-dependent indistinguishable error level for $\delta \phi (f)$  or $\delta A(f)$ for the three reference detector $S_n(f)$ and our reference neutron-star $A(f)$ at $d_\text{eff}=100\,$Mpc in Fig.~\ref{fig:goal-error}. These are translated into the shaded detector bounds in Figs.~\ref{fig:dph-compare-detector}.

Figure \ref{fig:goal-error} shows that waveform and calibration error of $\sim 3\%$  in amplitude (2 Deg of phase) through the tens of Hz range would be required for calibration to remain subdominant to stochastic noise for a single-detector $d_\text{eff}=100$ Mpc BNS with signal-to-noise ratio approximately 60 in the A+ detector. 

In Cosmic Explorer, the same system would have a signal-to-noise ratio of $\simeq$1158 above 10 Hz and require waveform and calibration errors as small as $10^{-3}$ for the impact of those uncertainties to be smaller than the statistical uncertainty.

\section{Range of waveform uncertainties for GW170817}

Recent discussions have raised the question of whether the analysis of gravitational-wave signals should be truncated at a specific frequency to limit systematic uncertainty in the inference of tidal properties, as those increase due to modeling differences at high frequencies (see for example \cite{Narikawa:2019xng,2021GambaBreschiBernuzzi}). We propose instead that a more appropriate treatment of increasing waveform uncertainty is to explicitly marginalize over the high-frequency $\delta\phi(f)$ range implied by those model differences. Analytic marginalization procedures or likelihood reweighting \cite{2022EssickCalibration,Payne_2019} could make this more tractable.

\begin{figure}
    \centering
  \includegraphics[width=0.49\textwidth]{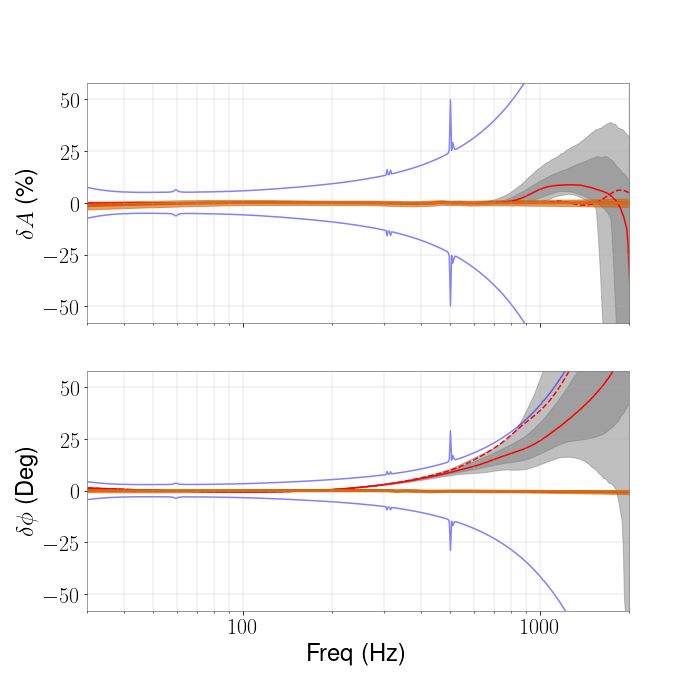}
\includegraphics[width=0.49\textwidth]{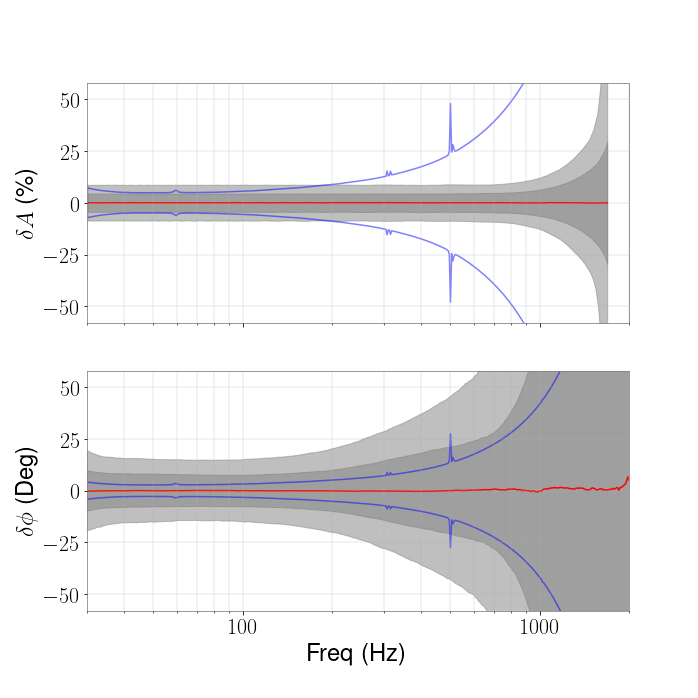}
    \caption{
    Left: waveform amplitude and phase differences over the posterior region of parameters in the \texttt{bilby} samples for GW170817 \cite{2020RomeroShawEt}. Differences are calculated between \texttt{IMRPhenomPv2\_NRTidalv2}\cite{Husa:2015iqa,2016KhanHusaHannamOhmePhenom,Dietrich:2017aum} and \texttt{SEOBNRv4T\_surrogate}\cite{Lackey:2018zvw} and projected onto LIGO Livingston (L1). The solid red line is the median distribution without spin terms, and 1-$\sigma$, and 2-$\sigma$ ranges around that median are shaded in grey. The dashed red lines show median waveform differences with spin but without tidal terms.
    The relatively small GWTC-1 envelopes for L1 calibration uncertainty during GW170817 \cite{P1900040-CALENV} are shown in orange.
    
    Right: to compare with realistic waveform uncertainties, the distribution of amplitude and phase differences at L1 between two random draws from the posterior distribution  is shown.  The PSD-based estimate of 1-$\sigma$ error from L1 noise fluctuations at the time of GW170817 \cite{P1900011-PSDs}, as seen in the shaded region of Figure \ref{fig:dph-compare-detector}, is outlined in blue in all panels.
}
    \label{fig:lowSpinPos}
\end{figure}

As a first estimate, we characterize the behavior $\delta A(f)$ and  $\delta\phi(f)$ for GW170817. Waveform modeling uncertainty is illustrated by generating an equivalent of the results of Fig.~\ref{fig:dph-compare-detector} 
for $\delta A(f)$ and $\delta\phi(f)$ between two waveform families (here, \texttt{IMRPhenomPv2\_NRTidalv2}\cite{Husa:2015iqa,2016KhanHusaHannamOhmePhenom,Dietrich:2017aum} and \texttt{SEOBNRv4T\_surrogate}\cite{Lackey:2018zvw}) at parameter values drawn from the the posterior distributions of the \texttt{bilby} samples for GW170817 \cite{2020RomeroShawEt}. Waveform amplitudes are calculated using the sampled distance, sky position and inclination, and projected on to the LIGO Livingston (L1) detector antenna pattern at the signal GPS time using a slightly modified version of \texttt{pesummary} \cite{HoyRaymond2021PESUMMARY}.  
Residual $\delta \phi$ is calculated as described in Section \ref{sec:waves}, by subtracting of the phase contribution of the maximum likelihood value for an overall shift in time and phase describing the waveform differences. To compare the effects of spin modeling and tidal modeling, we first calculate waveform differences with spin set to zero, and then with tides set to zero. The median $\delta A(f)$ and $\delta \phi(f)$ in both cases, and the distribution over the tidal posterior range, are shown in Figure \ref{fig:lowSpinPos}.  

Calibration envelopes for GW170817 from GWTC-1 \cite{P1900040-CALENV} are also shown on the same scale in Figure \ref{fig:lowSpinPos}. Calibration uncertainties are larger than waveform uncertainties at low frequencies, but both are small (sub 2 $\deg$ or 3\%) at low frequencies. In this comparison, waveform phase uncertainties become larger than calibration uncertainty starting between 100 and 200 Hz, while waveform amplitude uncertainties remain lower than calibration uncertainty up to 700 Hz.

To compare the waveform systematics to realistic uncertainties, we also show the distribution of $\delta A(f)$ and $\delta\phi(f)$ within the inferred posterior distribution of waveforms from the same \texttt{bilby} samples. Two waveforms are generated from two random parameter draws, both with the \texttt{IMRPhenomPv2\_NRTidalv2} model that was used for inference. The waveforms are again projected on to the L1 observatory. Here, the $t_c$ and $\phi_c$ parameters fix the full phase difference between the waveforms; each is generated with its sample's $\phi_c$, and the effect of $t_c$ on the Fourier-domain phase is incorporated by subtracting $2 \pi f \Delta t_c$ using the difference in coalescence time between the samples.

As a comparison, a PSD-based estimate of the 1-$\sigma$ uncertainty from statistical fluctuations is also shown, using the L1 noise curve \cite{P1800061-170817} and following the method of Section \ref{sec:requirements}. 
We note that the PSD-based estimates of 1-$\sigma$ waveform errors are an imperfect estimate of realistic uncertainties. The amplitude prediction shows reasonable agreement with the most constrained frequencies in the posterior. The waveform model has limited flexibility in amplitude before the merger, so it can not saturate the statistical range at all frequencies. However, phase differences within the posterior distribution have a 1-$\sigma$ level that is somewhat larger than the PSD-based estimate.  

The distribution of errors within the posterior distribution can also be compared to the credible intervals for $\delta A$ and $\delta \phi$ found in Edelman et al \cite{2021Edelman}. Their recovered-spline distribution for $\delta \phi$ is flatter over frequencies at the 1-$\sigma$ level, but at the 2-$\sigma$ level the credible intervals increase with increasing frequency (as seen in their Figures 5 and 12) in a similar way as the posterior- and PSD-based estimates shown here.

The waveform differences shown here come from a combination of different spin models, different tidal models, and differences rising from a re-summed effective-one-body waveform model vs. a frequency-domain phenomenological waveform generation. The range of tides and the presence of spin each affect the relative differences and their frequency dependence. The results confirm that systematic error for neutron-star inference can be reduced by restricting analysis to frequencies below $\sim1$kHz, as $\delta \phi$ are smaller than the statistical background in that range. 
They also confirm that amplitude uncertainty is not significant for GW170817.

However, this type of waveform comparison could be used to generate a distribution model for $\delta \phi$ and $\delta A$ that could be applied for coherent frequency-dependent waveform marginalization in proposed future event analyses. 
With this in mind, we computed the distribution of $\delta \phi$ and $\delta A$ over the full range of prior samples with the \texttt{IMRPhenomPv2NRT\_lowSpin\_prior} in the GW170817 PE release of GWTC-1 \cite{GWTC1}, using optimized versions of the waveform models \texttt{IMRPhenomPv2\_NRTidal\_INTERP} \cite{2022NitzPycbc} and \texttt{SEOBNRv4T\_surrogate} \cite{Lackey:2018zvw}. Even this low-spin prior range of parameters gives gives very wide distribution of both $\delta \phi (f) $ and $\delta A(f)$ at several hundred Hz; the 2-$\sigma$ range exceeds 50 Deg/\% between 200\,Hz and 1\,kHz. Most systems in the prior have poorly-defined contributions above $\sim1$\,kHz as the stars have merged at low frequencies due to large size and tides at larger values of the component $\Lambda$s. 

A realistic implementation of marginalization would therefore require a prescription for the distribution of $\delta \phi(f)$ that varies over the intrinsic parameters. Such a marginalization could also reflect the increase in modeling uncertainties away from well-calibrated regions of mass ratio.  While such a distribution could first be estimated by comparing multiple waveform models as is done here, it may also be better derived from theoretical estimates of intrinsic model uncertainties as discussed in Section \ref{sec:waves}.


\section{Astrophysical  interpretation of $\delta \phi$ and $\delta A$}
\label{sec:energetics}

Considering the size of the energy transfers required to modify the characteristic functions $\dot F (F)$ and $\mathcal{A}(F)$ will allow us to generally interpret constraints on $\delta \phi(f)$ and $\delta A(f)$, as might be recovered from a coherent analysis like that of Edelman et. al. \cite{2021Edelman}. 

Consider the usual energy balance equation for the evolution of a source system, with a total system energy $E(F)$ (for orbits: negative relative to $F\to0$) that can be written as a function of the instantaneous gravitational-wave frequency $F$.  
For example, two masses $m_1$ and $m_2$ in circular orbits emitting gravitational waves at $F = \Omega_{\text{orb}}/ \pi$ for orbital angular frequency $\Omega_{\text{orb}}$ have a leading order orbital binding energy of
\begin{equation}
\label{eq:totalen}
    E_\text{orb} = - \frac{m_1 m_2}{m_1+ m_2} \frac{c^2}{2}\left(\frac{G (m_1+m_2) (\pi F)}{c^3}\right)^{2/3}
\end{equation}
where the masses combine to a reduced mass and the term to the power of 2/3 is a dimensionless frequency parameter often denoted in post-Newtonian expansions as $x$ \citep{Blanchet2014LRR} .
The gravitational-wave energy loss or luminosity $\mathcal{L}_\text{GW}$ from the same orbit is
\begin{equation}
\mathcal{L}_\text{GW} = \frac{32}{5} \frac{c^5}{G} \frac{m_1^2 m_2^2}{\left(m_1+m_2\right)^4} \left(\frac{G (m_1+m_2) (\pi F)}{c^3}\right)^{10/3}
\end{equation}
using the same dimensionless frequency term.

Orbital evolution is driven by the energy balance at each characteristic emission frequency, which results in the  characteristic $\dot F$ function in terms of the energy gradient $E'(F) = d E(F)/ d F$ and the total rate of energy loss or luminosity $\mathcal{L}(F)$:
\begin{subequations}
\begin{align}
    \frac{d E(F)}{dt} &= - \mathcal{L}(F) &\text{and} &&
    \dot{F} &= - \frac{\mathcal{L}(F)}{E'(F)}
\end{align}
\end{subequations}

The gravitational-wave luminosity is determined by the multipole decomposition of strain emitted by the source \cite{2002BakerCampanelliLoustoTakahashi, 2020TaylorVarma}:
\begin{align}
    \mathcal{L}_\text{GW} 
    &= \frac{1}{4\pi} \int d\Omega \sum_{\ell m} Y_{-2}^{\ell m}(\Omega) 
    \left|d \dot h_{\ell m} (f) \right|^2 \\
    &= \frac{1}{16 \pi} \sum_{\ell m} d^2 \left( 
        \dot{\mathcal{A}}_{\ell m}^2 
        + (2 \pi)^2 \mathcal{A}_{\ell m}^2 F^2 
        \right)
\end{align}
where we use again $F =\dot{\psi} / 2\pi$ and where $d$ the luminosity distance at which $h$ is measured. 

When the SPA applies (following Eq.~\ref{eq:spa-conditions}) we can neglect the $\dot{\mathcal{A}^2}$ term, so gravitational-wave luminosity in each mode determines the waveform amplitude through 
\begin{equation}
    \mathcal{A}^2=  \frac{4}{\pi} \frac{  \mathcal{L}_\text{GW}}{d^2 F^2}
\end{equation}
The characteristic time-domain amplitude is directly connected to the total luminosity of the sytem.


The energy balance implied by a waveform model could be modified by any unmodeled physical effect which increases gravitational-wave luminosity $\mathcal{L}_\text{GW}(1+\delta \mathcal{L}_\text{GW})$, such as the coherent excess quadrupole moment induced by tides on each component neutron star.
Unmodeled non-gravitational-wave energy losses (including neutrino, electromagnetic,...) could also contribute to the system's evolution through an additional $\mathcal{L}_\text{tot}(1+\delta \mathcal{L}_\text{MM})$ that drives the energy balance without affecting the gravitational-wave amplitude.  
Finally, models derived from or calibrated to numerical relativity simulations might include unphysical excess energy loss due to numerical dissipation. The true waveform would then have $\mathcal{L}_\text{tot} (1 - \delta \mathcal{L}_\text{num})$; this is equivalent to a negative $\delta \mathcal{L}_\text{MM}$ needed to generate the true model. 

\subsection{Internal energy transfers: adiabatic vs. dynamic}
\label{ssec:internalenergy}
The waveform model may also neglect effects that change the system energy associated with a characteristic gravitational-wave emission frequency, such as  high-order corrections to the orbital binding energy or a missing energy reservoir. For example, there may be changes in the energy stored in a component neutron star's internal modes.

Consider some effect that changes total energy $E$ as a function of $F$, for example changing the leading order energy of Eq.~\ref{eq:totalen} to $E(1 +\delta E)$. A total derivative with respect to $F$ gives the correction to the energy gradient term in the energy balance equation, $d E /dF = E'$, as
   \begin{equation}
       E' (1 + \delta E') = E'\left(1+ \delta E + \frac{E}{E'}  \left(\delta E\right)' \right)
   \end{equation} 
Drawing from the extensive discussion of potential energy transfer impacts in the literature (such as in \cite{1999HoLai,2008FlanaganHinderer, Flanagan:2006sb,2016HindererTaracchiniFoucart}), we interpret the first term as a correction from adiabatic effects: those where the change in $\delta E$ with frequency is slow. In this case, noting each $\delta$ term here represents the size of \emph{relative} corrections to the original system $E'$ and $E$, the correction $\delta E'_A = \delta E_A$. For adiabatic tides, the correction $\delta E_A$ is negative \cite{2008FlanaganHinderer}: energy going into the stellar deformation implies a less negative total energy at a given frequency.
   
In contrast, the second term characterizes dynamic effects, coming from a rapid energy change. For example, consider a resonant transfer of energy $\Delta E_D$  from the orbit to a stellar mode, starting at a characteristic gravitational-wave frequency $F_0$ over a bandwidth $\Delta F$. Around $F=F_0$, we have a change in the gradient
\begin{align}
    E' (1+ \delta E') &\simeq E' - \frac{\Delta E_D}{\Delta F}\\
                    &= E'
                   \left( 1 - \frac{E}{E'} \frac{\delta E_D}{\Delta F} \right)
\end{align}
where we introduce the relative energy transfer $\delta E_D = \Delta E / E_\text{orb}$.

Define  
    the energy transfer timescale $ t_D = \Delta F / \dot F$
and  the adiabatic decay timescale or gravitational-wave emission timescale $t_\text{A} = | E / \mathcal{L}| \simeq 2 F / 3 \dot F  $.
  Using the energy balance relationship $E' = - \mathcal{L} / \dot F$,
 we find as has previously been shown \cite{Flanagan:2006sb} that the impact of a dynamic energy transfer compared to adiabatic energy transfer is amplified:
 \begin{equation}
     \delta E'_D = -\frac{E}{E'} \frac{\delta E_D}{\Delta F} 
     = \frac{E}{\mathcal{L}}\frac{\dot F}{\Delta F}  \delta E_D
     \simeq \frac{3}{2} \frac{t_A}{t_D}  \delta{E_D}
    \label{eq:dynen}
 \end{equation}
Generally, then, internal energy transfers lead to a gradient term
\begin{equation}
     \delta E'  \sim  \delta{E_A} + \frac{t_A}{t_D}  \delta{E_D}
 \end{equation}
where $\delta{E_A} E_\text{orb}$ and  $\delta{E_D} E_\text{orb}$ characterize the amounts of energy transferred adiabatically or dynamically.

\subsection{Signal implications}
Assume that all unmodeled effects are very small corrections to the baseline energy and luminosity, so that their effects can be linearized. We find the impact on the characteristic functions
\begin{subequations}
\begin{align}
    \mathcal{A}_\text{true} &= \mathcal{A}_\text{model} \left( 1 + \delta \mathcal{L}_\text{GW}\right)\\
    \dot F_\text{true} &= \dot F_\text{model}\left( 1 - \delta E'  + \delta \mathcal{L}_\text{GW} + \delta \mathcal{L}_\text{MM}\right) \\
    T'_\text{true} &= T' 
        + T' \left(\delta E' - \delta \mathcal{L}_\text{GW} - \delta \mathcal{L}_\text{MM}\right)
    \end{align}
\end{subequations}
This is consistent with tidally driven increases in the $\dot F$ function that can be seen at moderate frequencies for all binary neutron star models relative to the binary black hole model in Fig.~\ref{fig:FFdotf}.

To confirm the interpretation of these energy transfers, we work out the impact on the time-domain waveform following Eqs.~\ref{eq:Tpsidef} and using the energy balance relation  $T' =- E' /\mathcal{L}$,
\begin{subequations}
\begin{align}
    \delta T(f) &= \int_f^{f_c} dF  \frac{E'}{\mathcal{L}} \left(- \delta E' +
    \delta \mathcal{L}_\text{GW} + \delta \mathcal{L}_\text{MM} 
    \right)
    \\
    \delta\psi(f) &=  2 \pi \int_f^{f_c} dF  F  \frac{E'}{\mathcal{L}}  \left(-\delta E' +
    \delta \mathcal{L}_\text{GW} + \delta \mathcal{L}_\text{MM} 
    \right)
\end{align}
\end{subequations}
To interpret them physically: since orbital $E'<0$, this shows that both the time to coalescence and the number of cycles before coalescence can be shortened by additional luminosity or by a less rapid decrease of total energy of the system with increasing $F$. 

After propagation through to the signal domain, again linearizing the different corrections, we will find from Eqs.~\ref{eq:signal-from-model-uncertainty} the energetics implications for the frequency-domain signal: 
\begin{subequations}
\begin{align}
\delta A(f) &= \frac{1}{2} \left( \delta E' - \delta \mathcal{L}_\text{GW} - \delta \mathcal{L}_\text{MM}\right) + \delta \mathcal{L} _\text{GW} \\
        &= \frac{1}{2} \left( \delta E'+ \delta \mathcal{L}_\text{GW} 
        - \delta \mathcal{L}_\text{MM}\right)
        \\
\delta \phi(f) &= 2\pi f \delta T(f) - \delta \psi(f) \label{eq:dphi-en}\\
         &= - 2 \pi  \int_f^{f_c} d \tilde{f}\int_{\tilde{f}}^{f_c} dF \,\frac{E'} {\mathcal{L} }  \left( \delta E' - \delta \mathcal{L}_\text{GW} - \delta \mathcal{L}_\text{MM}\right)
\end{align}
\end{subequations}
Note that any energy losses that are not in gravitational waves lead to a decrease in gravitational-wave amplitude at the corresponding frequency - the orbit sweeps through more quickly and fewer cycles contribute. Phase accumulates more rapidly with any additional energy losses from the system.  The relative change in $\delta A$ vs $\delta \phi$ depends on the type of energy transfer that is added to the source model.

\subsection{Example: dynamic energy transfer and GW170817}
Applying the energetics framework above allows the interpretation of Edelman et al \cite{2021Edelman} results for GW170817 in terms of a possible augmentation of the inspiral waveform due to neutron-star energy transfers.  While no confident identification of a departure from the signal model was identified, the posterior distribution of $\delta \phi(f)$ in Figure~12 in \cite{2021Edelman} shows a 1-$\sigma$ excess at around $60$\,Hz of perhaps 5 Deg or order 0.1 radians.  At the same frequency, no significant $\delta A$ change is observed.

One candidate for a short-duration $\delta \phi(f)$ at a specific inspiral frequency is dynamic energy transfer into a neutron-star mode, for example as explored in \cite{2018AnderssonHo,2012TsangRead}. As the net impact on the Fourier-domain phase comes from a difference between $ 2 \pi f \delta T$ and $ \delta \phi$ in Eq.~\ref{eq:dphi-en}, a completely instantaneous energy loss only affects amplitude at the specific emission frequency. Consider instead the transfer of $\Delta E$ at $F_0$ over a small range $\Delta F$. If the excitation does not induce any coherent quadrupole,
we have a corresponding short-duration relative decrease in the signal amplitude from Eq.~\ref{eq:dynen} of
\begin{equation}
    \delta A(F_0) = \frac{1}{2} \delta E' 
    = \frac{1}{2}\frac{\Delta E}{t_D} \frac{1}{\mathcal L}
    = \frac{1}{2}\frac{t_A}{t_D} \delta E_D
\end{equation}
using the timescale notation of Sec.~\ref{ssec:internalenergy}.
If amplitude variation is not observed, this limits the total unmodeled energy transfer rate to be small compared to the luminosity at 60\,Hz, which is $\sim 2\times 10^{-4} M_\odot c^2 $/s or $-3 \times 10^{43}$ joules/s at 60 Hz. 
At the same time, there is an increase in accumulated Fourier-domain phase $\delta \phi$ which will be nonzero for frequencies between $F_0$ and $F_0+\Delta F$. We estimate the nonzero values as
\begin{align}
    \delta \phi 
    &\simeq 2 \pi \delta E' \int_{F_0}^{F_0+\Delta F} d \tilde{F}\int_{\tilde{F}}^{F_0+\Delta F} dF \, T'(F)  \\
    &\simeq \pi \delta E' \, \frac{ \left(\Delta F\right)^2}{\dot{F}}
    \simeq \pi  \Delta E \frac{\dot{F}}{\mathcal{L}}  t_D
    = \frac{2\pi}{3} t_D F \delta E_D
\end{align}
where $t_D F$ is the number of cycles over which the energy is transferred. 

A resonant energy transfer has a dynamic timescale of $t_D \sim \sqrt{t_A/F}$ \cite{Flanagan:2006sb}, so that both $ t_A / t_D$ and $ t_D F \sim \sqrt{ t_A F}$. At 60 Hz, $\sqrt{ t_A F} \simeq 40$, and the orbital binding energy $E \simeq -0.006 M_\odot c^2$ or $-1.1\times10^{45}$ joules, and a  energy transfer of $\delta E = \Delta E / E \sim 0.001$ at 60 Hz corresponds to roughly $\delta A = 2 \%$ and $\delta \phi = 5$ Deg, compatible with the credible intervals recovered for GW170817 \cite{2021Edelman}. We leave more realistic explorations of astrophysical interpretations to future work.


\section{Conclusion}

An amplitude-phase decomposition of time-domain gravitational-wave emission enables the use of instantaneous frequency in characterizing the underlying model physics in terms of characteristic functions for the signal evolution $\dot{F}$ and the time-domain amplitude $\mathcal{A}$. Resulting model uncertainties can be propagated from the time-domain wave to the frequency-domain signal for comparison with gravitational-wave observations. The differences in model waveform predictions that propagate from the underlying physics can be decoupled from the time and phase constants $t_c$ and $\phi_c$ that characterize a specific observation by finding maximum-likelihood values. Writing waveform uncertainties in the same form as calibration uncertainties --- namely, as frequency-dependent signal amplitude and phase error terms $\delta A(f)$ and $\delta \phi(f)$ --- would then characterize model-dependent waveform uncertainty suitable for marginalization in gravitational-wave inference.

Especially for observations of early inspiral, energy-balance arguments can aid in the interpretation of coherent waveform deviations for binary signals. We have illustrated how a common astrophysical energy transfer scenarios can be used to interpret recovered values of $\delta \phi(f)$ and $\delta A(f)$ compared to a waveform model used for analysis; recovered bounds on these functions limit the size of unmodeled energy transfers in the source system.

\section{Acknowledgements}
Many thanks to the IGWN Conda Distribution, numpy \cite{harris2020array}, pandas \cite{reback2020pandas}, watpy \cite{2022GonzalezZappaBreschiBernuzzi}, lalsimulation \cite{lalsuite}, pycbc \cite{2022NitzPycbc}, bilby \cite{2019AshtonBilby}, pesummary \cite{HoyRaymond2021PESUMMARY}, and the work of everyone that supports computational infrastructure for the LIGO-Virgo-Kagra collaborations and the gravitational-wave community. I thank David Radice, Wynn Ho, Sanjay Reddy, Josh Smith, Nils Andersson and the Caltech LIGO group for  helpful discussion. Thanks also to Aaron Zimmerman for useful and detailed feedback and corrections to Section 7, and the anonymous reviewers for especially helpful suggestions for Section 6. Read was supported during the development of these ideas by funding from NSF PHY-1307545 and PHY-1806962, the LIGO Lab, the Carnegie Observatories, and the Nicholas and Lee Begovich Center for Gravitational-Wave Astronomy. This material is based upon work supported by NSF's LIGO Laboratory which is a major facility fully funded by the National Science Foundation.

\appendix

\bibliography{refs}
\end{document}